\documentstyle[twoside, epsfig]{article}

\input ibvs2.sty

\begin{document}
\IBVShead{5975}{14 February 2011}

\IBVStitle{A 116 Year Record of Mass Transfer in R Arae}

\IBVSauth{Reed, Phillip A. }

\IBVSinst{Kutztown University, Kutztown, Pennsylvania, USA, e-mail: preed@kutztown.edu}

\SIMBADobjAlias{R Ara}{HD 149730}
\IBVStyp{CWA}
\IBVSkey{photometry}
\IBVSabs{This is a period study of the bright interacting southern eclipsing binary}
\IBVSabs{star R Arae.  New photometric data are combined with archival data to}
\IBVSabs{determine R Ara's average mass transfer rate over the past 116 years through}
\IBVSabs{the analysis of the resulting ephemeris curve.  An updated ephemeris is given.}

\begintext

R Arae (HD 149730) is a bright interacting southern binary star consisting of a B9 primary and a yet unseen secondary, undergoing rapid mass transfer just past the reversal of mass ratio stage of its evolution.  With an orbital period of 4.4 days, its components are close enough to experience a direct impact of mass transferring from the secondary to the primary, but distant enough that an accretion structure has formed around the primary.  The intense variations seen both photometrically and spectroscopically indicate that the accretion structure is unstable and quite variable (Reed {\it et al.}, 2010).  Because of this, R Ara is of great interest to the study of the evolution of interacting binary stars, but it has unfortunately been neglected.

New observations are combined with those found in the available literature (Hertzsprung, 1942; Payne-Gaposchkin, 1945; Nield, 1991; Reed, 2008 and Reed {\it et al.}, 2010) and in the database of the American Association of Variable Star Observers (AAVSO) to construct R Ara's first ephemeris curve, which plots observed-minus-calculated (O-C) times of primary eclipses and spans the 116 years since its discovery by Roberts (1894).  The best-fit to the O-C curve is a quadratic function with parameters that yield period change and mass transfer rates consistent with those of an active Algol-type interacting binary.

The new observations presented here were collected at the Tzec Maun Observatory, located near Moorook, South Australia.  The telescope is a 15.2-cm, f/7.3 refractor equipped with a research-grade CCD camera (3072x2048, 9-$\mu$ pixels).  Each image was exposed for 10 seconds through a Bessel-V filter.  The observations of 21 May 2010 consist of 51 consecutive images taken over eight hours, and those of 30 May 2010 consist of 50 consecutive images taken over five hours. The comparison stars were HD 150185 (HIP 81611), HD 149715 (HIP 81581), and HD 149784 (HIP 81611).  The comparisons were chosen due to their proximity to R Ara and the fact that they are known to not be variable themselves ($\sigma _V < 0.01$ mag).  The comparison stars' magnitudes are listed in the Hipparcos/Tycho archive.

All observed times of primary minimum are compiled in Table 1 and are plotted in Figure 1.  The plot is somewhat sparsely populated, as evidence of R Ara's neglectedness, but it clearly indicates true period change.  The observations marked with * in Table 1 refer to times of minimum light that were determined for this study, using the method of Kwee and VanWoerden (1956).  The ephemeris found by Nield (1991) of $HJD_{Pr.Min.} = 2446585.1597 + 4.425132 E$ was used to compute the calculated eclipse times.  The AAVSO data are plotted in Figure 2.

\begin{center}
\begin{table}
\caption{Observed times of primary minimum of R Ara.}
	\centering
		\begin{tabular}{cccc}
		\hline
		\textbf{HJD (Pr.Min.)} & \textbf{Date} & \textbf{Observer / Reference}\\
		\hline
2412954.373\phantom{000} & 05 May 1894 & AAVSO*\\
2412985.356\phantom{000} & 05 June 1894 & AAVSO*\\
2413016.320\phantom{000} & 06 July 1894 & AAVSO*\\
2413370.327\phantom{000} & 25 June 1895 & AAVSO*\\
2413755.311\phantom{000} & 14 July 1896 & AAVSO*\\
2414547.406\phantom{000} & 14 September 1898 & AAVSO*\\
2415140.355\phantom{000} & 30 April 1900 & AAVSO*\\
2416348.428\phantom{000} & 21 August 1903 & AAVSO*\\
2416963.461\phantom{000} & 27 April 1905 & AAVSO*\\
2417742.251\phantom{000} & 15 June 1907 & AAVSO*\\
2425818.028\phantom{000} & 25 July 1929 & Hertzsprung (1942)\\
2428402.28\phantom{0000} & 21 August 1936 & Hertzsprung (1942)\\
2429433.348\phantom{000} & 18 June 1939 & Payne-Gaposchkin (1945)\\
2446585.1597\phantom{00} & 03 June 1986 & Nield (1991)\\
2454501.932\phantom{000} & 05 February 2008 & Reed (2008)\\
2454541.757\phantom{000} & 16 March 2008 & Reed, \textit {et al.} (2010)\\
2455338.3037\phantom{00} & 21 May 2010 & Reed (this paper)*\\
2455347.1559\phantom{00} & 30 May 2010 & Reed (this paper)*\\

\hline
		\end{tabular}

\end{table}
\end{center}

\IBVSfig{10cm}{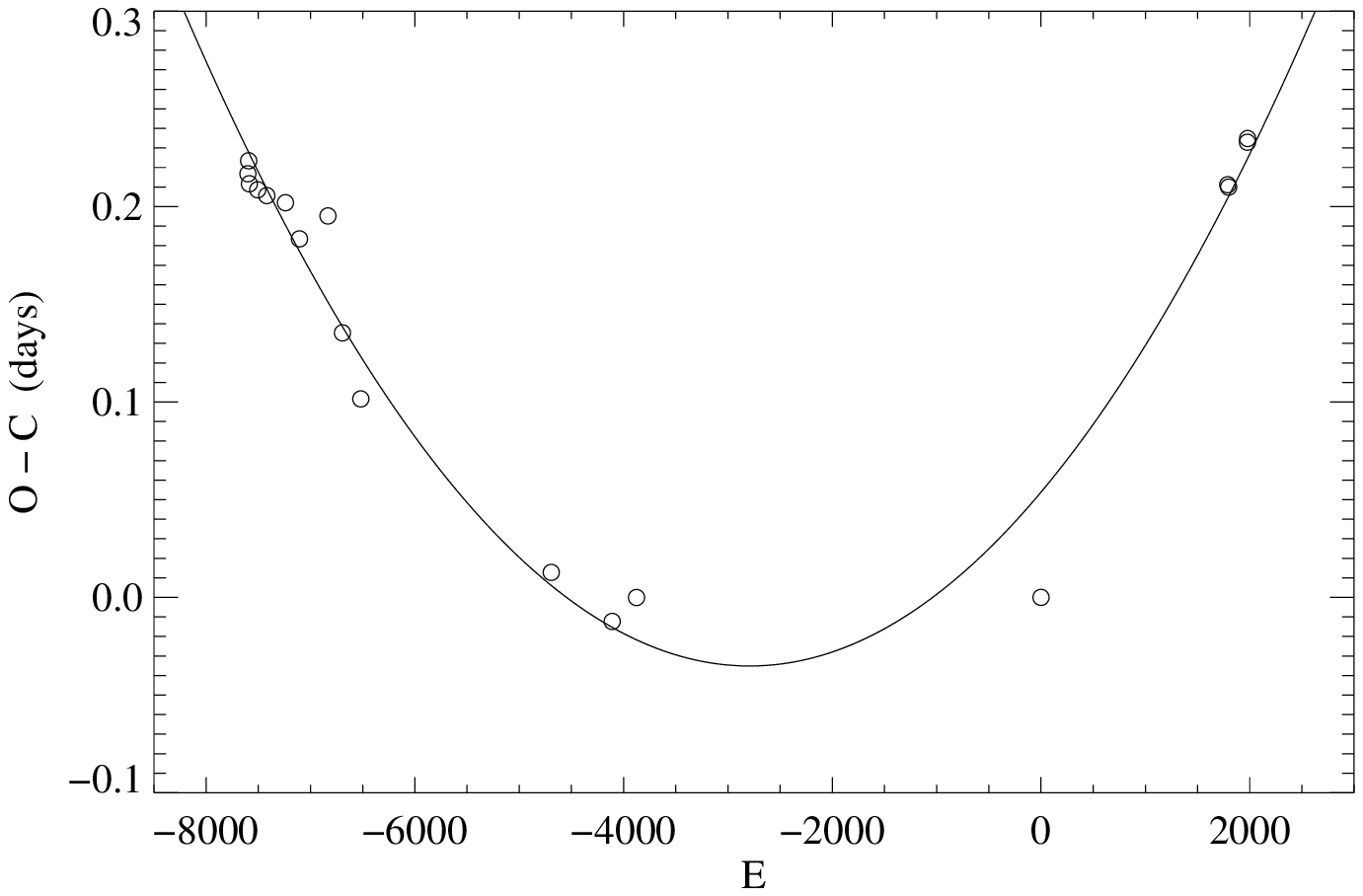}{
This is the first ephemeris (O-C) curve for R Ara, which spans 116 years.  The line is the quadratic function that is the best fit to the data points.}

\vskip 2mm

The best-fit O-C curve is given by:
\begin{center}
$ O-C = (0.0538) + (6.371 \times 10^{-5}) E + (1.141 \times 10^{-8}) E^2 $
\end{center}

The average rate of period change over the past 116 years is calculated to be:
\begin{center}
$ \dot{P} = \frac {2C_2}{P} = \frac{2(1.141 \times 10^{-8})}{4.425132} = 5.16 \times 10^{-9} \frac{days}{day} $
\end{center}

Then, using Sahade's values for the masses of the stars of $M_1=4M_{\odot}$ and $M_2=1.4M_{\odot}$ (Sahade 1952), and assuming conservative mass exchange, the rate of mass transfer averaged over the past 116 years is:
\begin{center}
$ \dot{M} = \frac {\dot{P} M_1 M_2} {3 P (M_1 - M_2)} = \frac {(5.16 \times 10^{-9})(4)(1.4)}{3(4.425132)(4-1.4)} = 8.37 \times 10^{-10} \frac {M_{\odot}}{day} $
\end{center}

or:
\begin{center}
$ \dot{M} = 3.06 \times 10^{-7} \frac {M_{\odot}}{year} $
\end{center}

which is consistent with an actively interacting Algol-type system undergoing rapid mass transfer.  Albright and Richards (1996) have stated that Algols transfer mass at rates ranging from $\sim10^{-11}M_{\odot}\ yr^{-1}$ to $\sim10^{-7}M_{\odot}\ yr^{-1}$. A system very similar to R Ara is U Sge, which was reported to exhibit a mass transfer rate of $\dot{M} \leq 2 \times 10^{-7}M_{\odot}\ yr^{-1}$ by Olson (1987) and $\dot{M} = 6.15 \times 10^{-7}M_{\odot}\ yr^{-1}$ by Manzoori (2008).  The timescale for this stage of R Ara's evolution is very short, on the order of 10,000 years.

The new times of primary minimum, which were determined to be at HJD2455338.303704 $\pm$ 0.000591 and HJD2455347.155868 $\pm$ 0.000378, provide an instantaneous orbital period of 4.426082 $\pm$ 0.000485 days. A new ephemeris of $ HJD_{Pr.Min.} = 2455338.303704 + 4.426082 E $ will provide more accurate calculated eclipse times, and other phase values, for future observations.  Figure 3 shows the May 2010 light curve of R Ara near primary minimum.

{\it Acknowledgements:}\\
I would like to thank the Tzec Maun Foundation for the use of their observatory, and the AAVSO for the valuable archival data available in their database.  I would also like to thank the referee for useful comments on this manuscript.

\references
Albright, G. E., and Richards, M. T., 1996, {\it ApJ}, {\bf459}, L99

Hertzsprung, E., 1942, {\it Bull.Astron.Inst.Neth.}, {\bf9}, 277

Kwee, K. K., and Van Woerden H., 1956, {\it Bull.Astron.Inst.Neth.}, {\bf 12}, 327

Manzoori, M., 2008, {\it Ap\&SS}, {\bf 313}, 339

Nield, K. M., 1991, {\it Ap\&SS}, {\bf 180}, 233

Olson, E. C., 1987, {\it AJ}, {\bf 94}, 1043

Payne-Gaposchkin, C. H., 1945, {\it Harvard Ann.}, {\bf 115}, 46

Reed, P. A., {\it et al.}, 2010, {\it MNRAS}, {\bf 401}, 913

Reed, P. A., 2008, PhDT, Lehigh University

Roberts, A. W., 1894, {\it AJ}, {\bf 14}, 113

Sahade, J., 1952, {\it ApJ}, {\bf116}, 27

\endreferences

\IBVSfig{20cm}{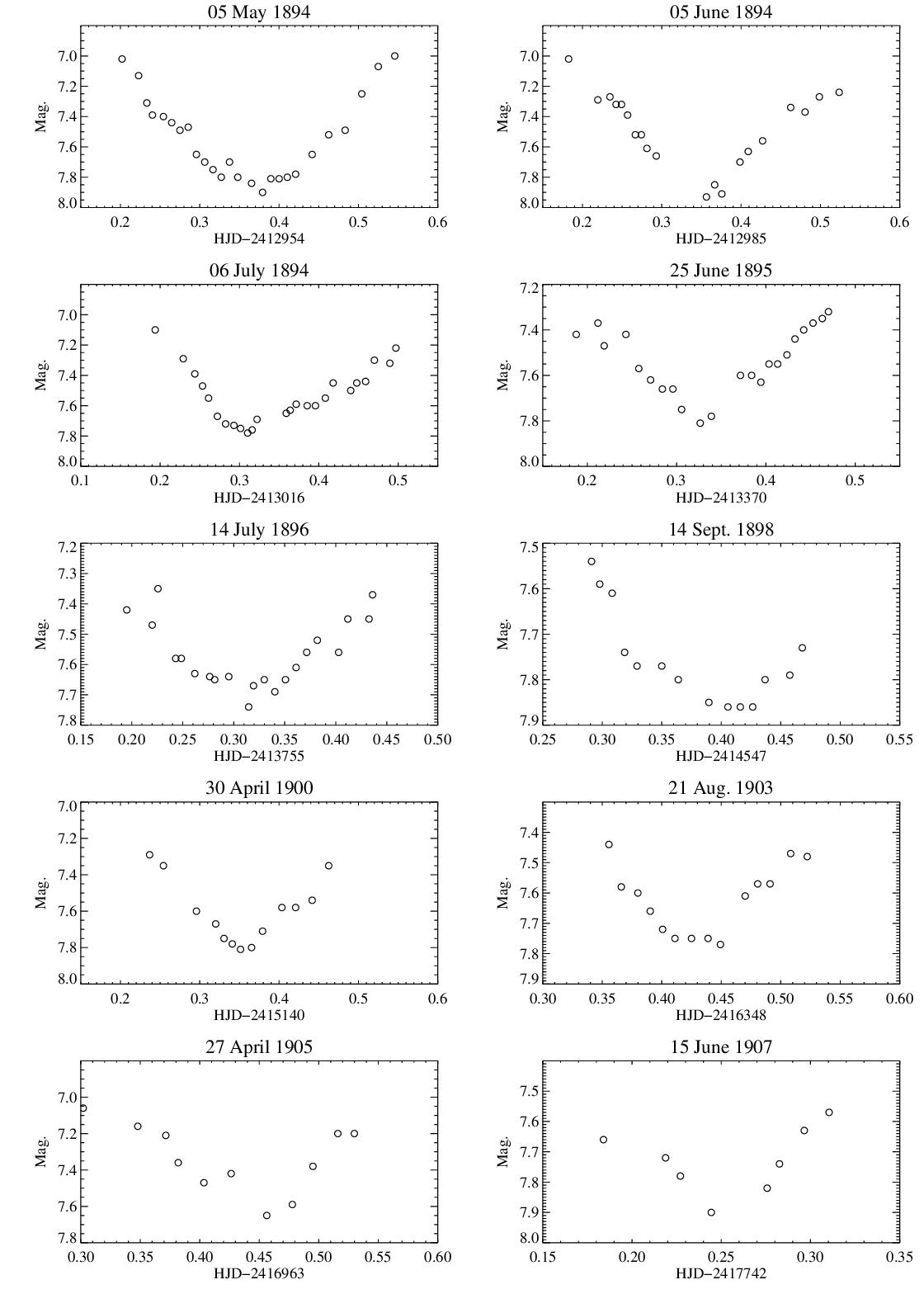}{
The archival AAVSO light curves.}

\IBVSfig{10cm}{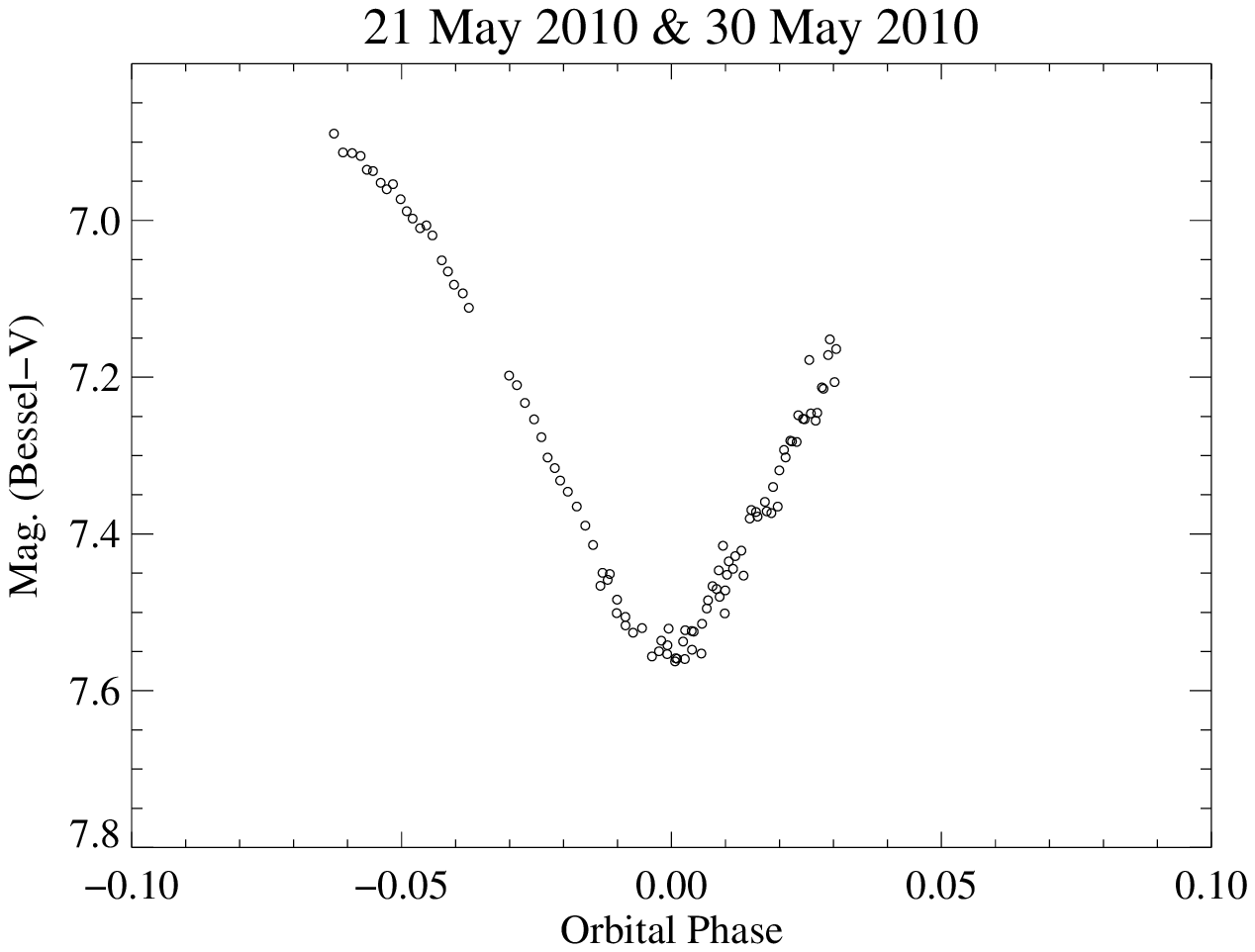}{
The 2010 light curve near primary minimum.  The orbital phase values were determined using the newly calculated ephemeris.}

\end{document}